\documentclass{PoS}

\title{Charmless B Decays at CDF}

\ShortTitle{Charmless B Decays at CDF}

\author{\speaker{Diego Tonelli}%
        \thanks{On behalf of the CDF Collaboration.}\\
       Fermilab, P.O. Box 500, Batavia, IL, 60510-5011,  USA\\
       E-mail: \email{tonel@fnal.gov}}

\abstract{The CDF experiment at the Fermilab Tevatron \pap\ collider pioneered the exploration of charmless \bhadron\ decays in hadron collisions. This provided a unique, rich and highly successful program of measurements that is currently reaching its full maturity. I report here results on two-body decays and the new analysis of $\bs \to \phi\phi$ decays in an event sample corresponding to 2.9~\lumifb.}

\FullConference{12th International Conference on B-Physics at Hadron Machines \\
		 September 7-11, 2009\\
		 Heidelberg, Germany.}

\def \rightdownarrow
 {\kern.3em
 \rule[.5ex]{.15mm}{2ex}
 {\mbox{$\kern-0.1em{\longrightarrow}$}}      % down-right arrow for cascade decay equations
 }

\def\lessim{\mathrel {\vcenter {\baselineskip 0pt \kern 0pt  % minore/circa uguale
\hbox{$<$} \kern 0pt \hbox{$\sim$} }}}

\def\gessim{\mathrel {\vcenter {\baselineskip 0pt \kern 0pt   % maggiore/circa uguale
\hbox{$>$} \kern 0pt \hbox{$\sim$} }}}

\newcommand{\eg}{e.~g.,}					%   	per esempio 

\newcommand{\lumipb}{\mbox{pb$^{-1}$}}				%	pb^-1
\newcommand{\lumifb}{\mbox{fb$^{-1}$}}				%	fb^-1

\newcommand{\mum}{\mbox{$\mu$m}}				%	um
\newcommand{\mus}{\mbox{$\mu$s}}

\newcommand{\br}{\ensuremath{\mathcal{B}}}

\newcommand{\tev}{\ensuremath{\mathrm{Te\kern -0.1em V}}}
\newcommand{\gev}{\ensuremath{\mathrm{Ge\kern -0.1em V}}}	%	GeV
\newcommand{\mev}{\ensuremath{\mathrm{Me\kern -0.1em V}}}	%	MeV
\newcommand{\kev}{\ensuremath{\mathrm{ke\kern -0.1em V}}}	%	keV
\newcommand{\massgev}{\mbox{\gev/$c^2$}}			%	GeV/c^2
\newcommand{\massmev}{\mbox{\mev/$c^2$}}			%	MeV/c^2
\newcommand{\pgev}{\mbox{\gev/$c$}}				%	GeV/c
\newcommand{\stat}{\ensuremath{\mathit{~(stat.)}}}		%	(stat.)
\newcommand{\syst}{\ensuremath{\mathit{~(syst.)}}}		%	(syst.)

\newcommand{\CP}{\ensuremath{\mathsf{CP}}}			%	CP

\newcommand{\pap}{\proton\antiproton}			%	ppbar
\newcommand{\pt}{\ensuremath{p_{\rm{T}}}}			%	pT

\newcommand{\lxy}{\ensuremath{L_{\rm{T}}}}

\newcommand{\proton}{\ensuremath{\mathrm{p}}}
\newcommand{\antiproton}{\ensuremath{\bar{\rm{p}}}}

\newcommand{\bd}{\ensuremath{B^{0}}}				% 	B-zero
\newcommand{\bs}{\ensuremath{B^{0}_s}}				% 	B (zero) sub-s

\newcommand{\bu}{\ensuremath{B^{+}}}				%	B piu'
\newcommand{\bhadron}{\mbox{$b$--hadron}}			%	b-hadron
\newcommand{\bn}{\ensuremath{B^{0}_{(s)}}}			% 	neutral B (B-zero and B sub-s)
\newcommand{\lambdab}{\ensuremath{\Lambda^0_{b}}}

\newcommand{\bhh}{\ensuremath{\bn \to h^{+}h^{'-}}}

\newcommand{\bdpipi}{\ensuremath{\bd \to \pi^+ \pi^-}}

\newcommand{\bdkpi}{\ensuremath{\bd \to K^+ \pi^-}}

\newcommand{\bskpi}{\ensuremath{\bs \to K^- \pi^+}}

\newcommand{\bskk}{\ensuremath{\bs \to  K^+ K^-}}

\newcommand{\bspipi}{\ensuremath{\bs \to  \pi^+ \pi^-}}

\newcommand{\bdkk}{\ensuremath{\bd \to  K^+ K^-}}

\newcommand{\bsjpsiphi}{\ensuremath{\bs \to  \jpsi \phi}}

\newcommand{\lambdabppi}{\ensuremath{\lambdab \to \proton \pi^{-}}}
\newcommand{\lambdabpk}{\ensuremath{\lambdab \to \proton K^{-}}}

\newcommand{\jpsi}{\ensuremath{J/\psi}}

\newcommand{\fig}[1]{fig.~\ref{fig:#1}}

\newcommand{\cita}[1]{\cite{#1}}

\newcommand{\dedx}{\ensuremath{\mathit{dE/dx}}}

\newcommand{\acp}{\ensuremath{A_{\mathsf{CP}}}}

\newcommand{\fracpipisukpi}{\ensuremath{\frac{\br(\bdpipi)}{\br(\bdkpi)}}}
\newcommand{\frackksukpi}{\ensuremath{\frac{\mathit{f_s}}{\mathit{f_d}}\times\frac{\br(\bskk)}{\br(\bdkpi)}}}

\newcommand{\fracdkksukpi}{\ensuremath{\frac{\br(\bdkk)}{\br(\bdkpi)}}}
\newcommand{\fracspipisukpi}{\ensuremath{\frac{\mathit{f_s}}{\mathit{f_d}}\times\frac{\br(\bspipi)}{\br(\bdkpi)}}}
\newcommand{\fracskpisukpi}{\ensuremath{\frac{\mathit{f_s}}{\mathit{f_d}}\times\frac{\br(\bskpi)}{\br(\bdkpi)}}}
\newcommand{\fracpksukpi}{\ensuremath{\frac{\mathit{f_L}}{\mathit{f_d}}\times\frac{\br(\lambdabpk)}{\br(\bdkpi)}}}
\newcommand{\fracppisukpi}{\ensuremath{\frac{\mathit{f_L}}{\mathit{f_d}}\times\frac{\br(\lambdabppi)}{\br(\bdkpi)}}}

\def\babar{\mbox{\slshape B\kern-0.1em{\smaller A}\kern-0.1em B\kern-0.1em{\smaller A\kern-0.2em R}}}
\newcommand{\belle}{Belle}

\begin{document}
\section{Introduction}
Non-leptonic charmless decays of $b$--hadrons are among the most widely studied processes in quark-flavor physics.  The Collider Detector at Fermilab (CDF) is the first experiment that explores these decays in hadron collisions, and has so far unique access to large samples of charmless \bs\  (and \lambdab) decays, necessary supplements to the wealth of precision measurements in the \bd\ and \bu\ sectors available from the $B$-factories \cite{ckm-review}. This is made possible by identifying long-lived decays in the trigger, where dedicated custom electronics \cite{svt} reconstructs accurately trajectories of charged particles (tracks) in the micro-vertex silicon detector. Tracks in the plane transverse to the beam are reconstructed at the first trigger stage (synchronous with the 2.52 MHz bunch-crossing) in the multi-wire drift chamber (that covers $|\eta|<1$)  and matched at the second trigger stage (asynchronous with $25~\mus$ average latency) with tracks found in the silicon detector. 
The innermost silicon layer, at 2.5 cm radius from the beam, allow measurement of impact parameter (distance of closest approach to the beam) with offline-like, 48~\mum\ resolution (including the beam spread). This is small with respect to $150~\mum$  values typical of relativistic $b$--hadron decays, allowing discrimination of long-lived heavy flavor decays from the overwhelming light-quark background.  The resulting samples of hadronic $B$ decays provided a rich and fruitful program that is currently reaching its maturity, and whose latest results are shown in what follows. All conjugate modes are implied, branching fractions are \CP-averages, and $K^*$ is shorthand for $K^*(892)^0$.
\section{Updated analysis of  $\bs\to\phi\phi$ decays}
The $\bs\to\phi\phi$ decay is mediated by a penguin-dominated $b\to s\bar{s}s$ transition and was first detected in year 2005 by CDF, with 8 events in a data sample corresponding to a time-integrated luminosity of 180~\lumipb\ \cita{bsphiphi}. The decay-rate is predicted with large uncertainty and the phenomenology is enriched by the presence of three different polarization amplitudes, which enhance sensitivity to non--standard-model (non-SM) physics.  We improved the measurement of branching fraction using 2.9~\lumifb\ of data.  The $\bs\to\phi(\to K^+K^-)\phi(\to K^+K^-)$ rate is measured using $\bs\to\jpsi(\to \mu^+\mu^-)\phi(\to K^+K^-)$ decays, collected by the same trigger, as a reference mode as follows:
\begin{equation}
\frac{\br(\bs\to\phi\phi)}{\br(\bsjpsiphi)} = \frac{N_{\phi\phi}}{N_{\jpsi\phi}}\frac{\br(\jpsi\to\mu^+\mu^-)}{\br(\phi\to K^+K^-)}\frac{\varepsilon_{\jpsi\phi}}{\varepsilon_{\phi\phi}}\varepsilon_{\mu},
\end{equation}
where $N_{\phi\phi}$ ($N_{\jpsi\phi}$) is the number of signal (reference) decays determined in data; the branching ratios of the vector mesons are  known \cita{pdg}; the trigger and selection efficiency to reconstruct a signal decay relative to a reference one,  $\varepsilon_{\phi\phi}/\varepsilon_{\jpsi\phi}$,   is determined from simulation; and the efficiency for requiring muon identification on one of the $\jpsi$ prongs is accounted for separately by the $\varepsilon_{\mu}$ term, determined from inclusive \jpsi\ in data. We chose \bsjpsiphi\ as a reference because it involves a $\phi$ decay,  and has same number of vertices and final state particles as the signal. The decay $\bd \to \phi(\to K^+K^-)K^*(\to K^+\pi^-)$ is even more similar to the signal, but its use as reference carries the penalty of the large uncertainty on the ratio of production fractions $f_s/f_d$. \par The trigger requires  two oppositely-charged particles originated in a point displaced transversely by the primary \pap\ collision vertex. Trigger tracks are required to have an impact parameter of $0.012<d_0<0.1$ cm, to originate  at least 200~\mum\ from the primary vertex in the plane transverse to the beam, and to be azimuthally separated by $2^\circ < \Delta\phi< 90^\circ$. Different thresholds on transverse momenta of particles and their scalar sum ($\pt>2~\mbox{or}~2.5~\pgev$, $\sum \pt > 5.5~\mbox{or} ~6.5~\pgev$) are applied as a function of luminosity to control trigger rates. %The resulting event sample contains $\bs\to\phi\phi$ candidates where typically both tracks that fire the trigger are kaons from the same $\phi$ meson. 
In the offline analysis, a kinematic fit to a common space-point is applied to each track pair. Pairs consistent with being kaons originated from $\phi$ meson decays are then fit together to a 4-tracks vertex.\par  The mass distribution of the four kaons shows a large, featureless flat background. Samples of  simulated data (to characterize signal, $S$) and \bs\ mass sidebands (for background, $B$) are used in an unbiased optimization to extract a visible signal. We maximized, over a large set of configurations of kinematic and vertex quality requirements, the quantity $S/\sqrt{S+B}$. This is close to optimal for a branching fraction measurement, since $S/\sqrt{S+B} \propto1/\sigma_{\br}$, where $\sigma^2_{\br}$ is the variance of a rate in a counting experiment. The most discriminating observables are the transverse decay-length of the candidate, $\lxy(\bs)>330~\mum$,  and the transverse momentum of the lower--\pt\ kaon, $\pt(K)>0.7~\pgev$. Other used quantities include impact parameter of the highest--\pt\ vector meson, $d_0(\phi)>85~\mum$; impact parameter of the candidate, $d_0(\bs)<65~\mum$; and the fit-quality of the reconstructed secondary vertex, $\chi^2<17$.
\begin{figure}[!h]
\begin{center}
\includegraphics[width=0.475\textwidth]{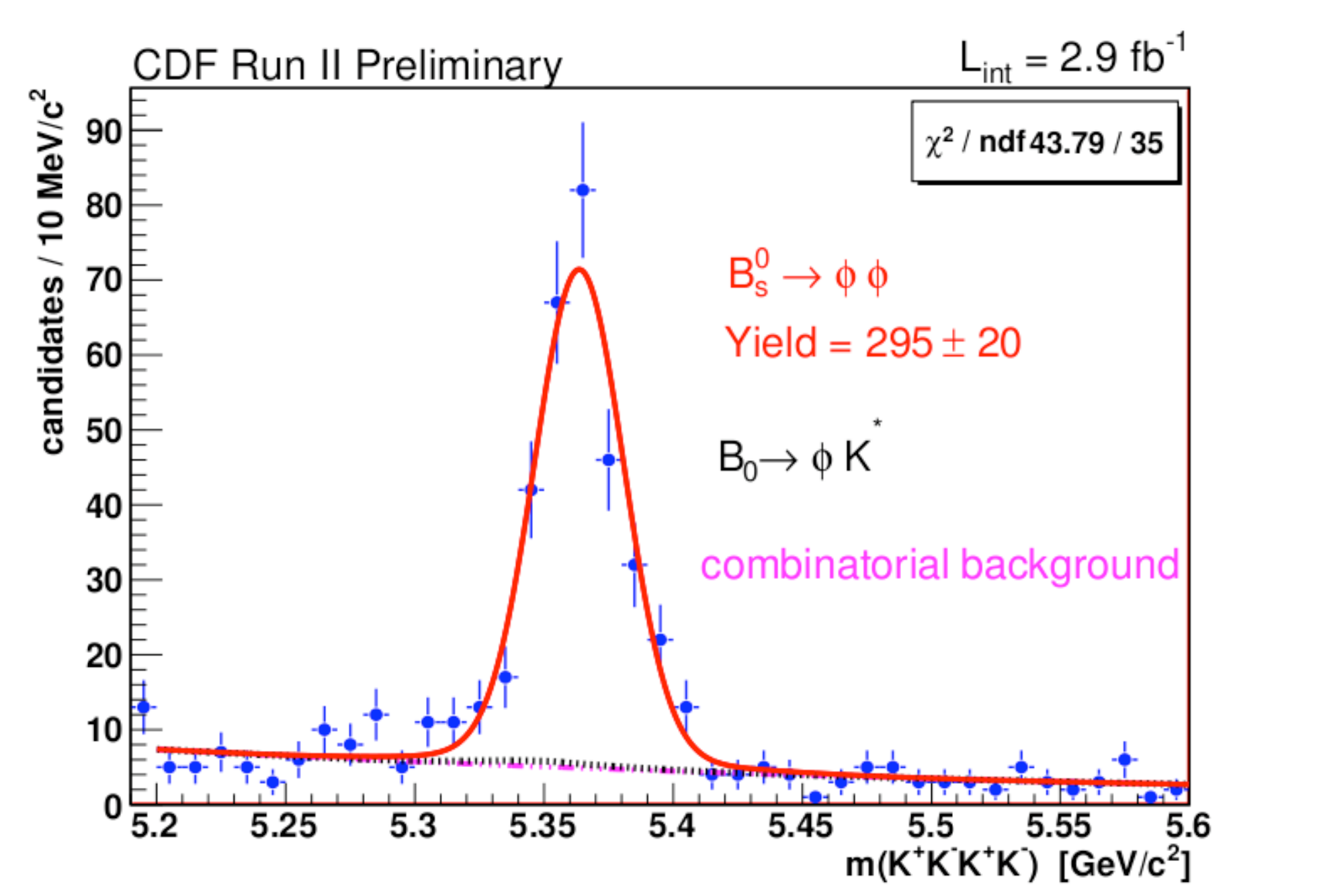}
\includegraphics[width=0.475\textwidth]{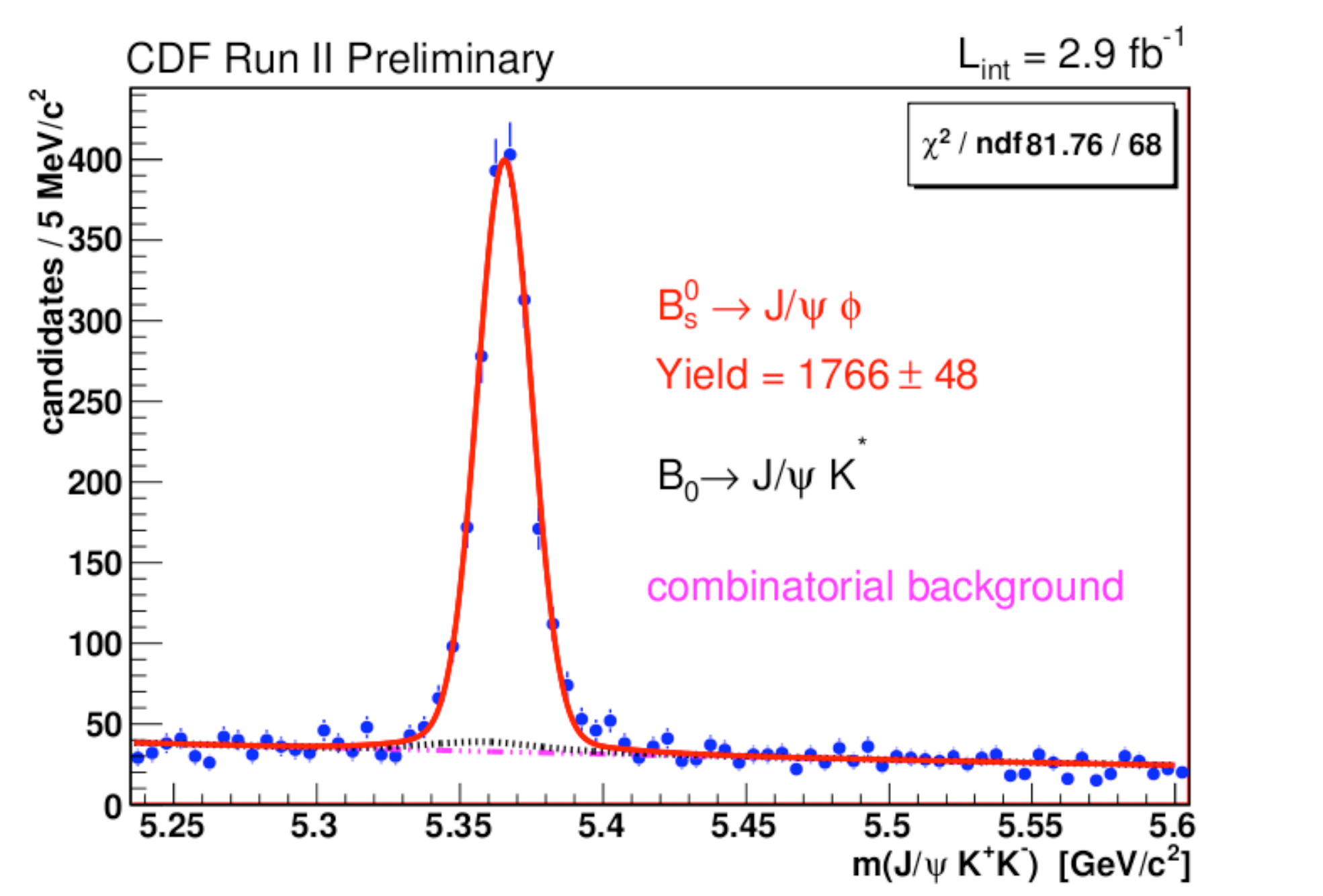}
\end{center}
\caption{Distribution of $K^+K^-K^+K^-$ mass (left) and  $\jpsi K^+K^-$ mass (right) with fit function overlaid.}
\label{fig:BVVmass}
\end{figure}
\par The final sample contains $295\pm20$ signal events over a moderate background (\fig{BVVmass}, left) dominated by accidental combinations of random tracks that meet the selection requirements (combinatorial). Simulation shows that a 2.5\% contamination of $\bd\to\phi K^*$ is also present and contributions of the (as yet unobserved) $\bs\to K^*K^*$ decay are negligible for branching fractions $\mathcal{O}(10^{-5})$, as predicted by the SM. The events clustering at masses of 5.25--5.30~\massgev\  suggest a possible $\bd\to\phi\phi$ contribution. A fit where a resonance at the \bd\ mass is allowed to float yields results consistent with a background fluctuation of less than $2\sigma$, with a corresponding upper limit on $\br(\bd\to\phi\phi)$ consistent with currently known values. This has been included as a systematic effect in the $\br(\bs\to\phi\phi)$ measurement. The factor 40 gain in signal yield with respect to the previous result \cita{bsphiphi}, in spite of just a factor 16 increase in sample size quantifies the effect of the optimized trigger and offline selection. Seven layers of silicon sensors extending radially up to 22 cm, and the drift chamber that provides 96 measurements between 30 and 140 cm, all immersed in the 1.4 T axial magnetic field, provide a mass resolution of approximately 17 \massmev\ on the $\bs\to\phi\phi$ peak.\par
Offline reconstruction of \bsjpsiphi\ decays is similar to the signal reconstruction, except that the 4-track vertex is fit using the known \jpsi\ mass as a constraint  and at least one track from the candidate \jpsi\ is required to have a matching extrapolation outward to track-segments reconstructed in the muon detectors (planar drift chambers at $|\eta|<1$). This requirement selects a sufficiently abundant and pure sample rejecting any $\jpsi\to e^+e^-$ contamination. A dedicated optimization yields $1766\pm48$ \bsjpsiphi\ events over a low background dominated by combinatorics (\fig{BVVmass}, right) and a residual 4\% contribution from mis-reconstructed $\bd\to\jpsi K^*$ decays. Note that  \bsjpsiphi\ decays reconstructed from the displaced-track trigger provide an additional 25\% signal for the analysis of the \bs\--mixing phase, which suffers from low-statistics issues \cita{sin2betas}.\par
The relative trigger and selection efficiency is extracted from simulation. Simulated decays were reweighed to match the transverse momentum distribution of data because the Monte Carlo does not reproduce accurately the observed admixture of triggers with different \pt-thresholds. The result is $\varepsilon_{\jpsi\phi}/\varepsilon_{\phi\phi} = 0.939 \pm 0.009 \pm 0.030$. The first uncertainty is due to finite statistics collected by each trigger,  the second one arises from the statistical uncertainty in the reweighing.\par
The efficiency for identifying a muon is extracted from data using a few thousands inclusive $\jpsi\to\mu^+\mu^-$ decays  as a function of muon \pt\ and in two ranges of pseudorapidity, $|\eta|<0.6$ and $0.6<|\eta|<1.0$, corresponding to distinct muon detectors. The \jpsi\ decays are collected by the same trigger that collects signal decays, which is unbiased for muons,  and selected offline as follows: $\lxy(\jpsi)>200~\mum$, $\chi^2 < 5$, $\pt(\mu)>1.5~\pgev$, and at least one \jpsi\--track that extrapolates outward to a track segment in the muon chambers. The resulting \pt--averaged efficiency is $\varepsilon_{\mu} = N_{2\mu}/(N_{1\mu} + N_{2\mu}) = 0.8695 \pm 0.0044 \stat $, where $N_{1\mu} (N_{2\mu})$ are \jpsi\ candidates that have exactly one (two) muon(s) identified offline, and the efficiency for identifying di-muons is assumed factorizable into the product of  single muon efficiencies. %Curves of \pt-dependent muon efficiencies are shown in \fig{MuonEfficiency} with empiric fit functions (yielding fit probabilities greater than 30\%) overlaid. 
\par
The dominant systematic uncertainty arises from the unknown polarizations of signal and reference decays (6-7\%), which impact acceptance. Other contributions include a $K/\mu$ difference in trigger efficiency due to different ionization in the tracking chamber (3-4\%),  uncertainty in signal mass model (3\%), unmodeled backgrounds (3\%), \pt-reweighing and background subtraction (1\%). From the  observed number of events, 
$N_{\phi\phi} = 295 \pm 20~\stat \pm 12~\syst$ and $N_{\jpsi\phi} = 1766 \pm 48~\stat \pm 41~\syst$, we determine
\begin{equation}
\frac{\br(\bs\to\phi\phi)}{\br(\bsjpsiphi)} = [1.78 \pm 0.14\stat \pm 0.20\syst]\times 10^{-2}
\end{equation} 
%\begin{equation}
%\br(\bs\to\phi\phi)/\br(\bsjpsiphi) = [1.78 \pm 0.14\stat \pm 0.20\syst]\times 10^{-2}
%\end{equation} 
which, using $\br(\bsjpsiphi) = (1.45\pm0.46)\times 10^{-3}$ (the known value updated with the current value of $f_s/f_d$ \cita{pdg}),  yields $\br(\bs\to\phi\phi) = [2.40 \pm 0.21\stat \pm 0.27\syst \pm 0.82 (\br)] \times 10^{-5}$. The last uncertainty derives from the uncertainty in \br(\bsjpsiphi). The result agrees with the previous determination, [$1.40 ^{+0.6}_{-0.5}\stat \pm 0.6\syst]\times 10^{-5}$ \cita{bsphiphi}, and with predictions based on QCD factorization,  $[2.18 \pm 0.1 ^{+3.04}_{-1.78}]\times 10^{-5}$  or perturbative QCD, $[3.53 ^{+0.83 +1.67}_{-0.69 - 1.02}]\times 10^{-5}$  \cita{theoryBRphiphi}, all of which have larger uncertainties. \par
This sample of  300 events is already sufficient to explore the rich polarization structure of the $\bs\to\phi\phi$ final state. Three states of polarization correspond to the allowed values of orbital angular momentum ($\ell = 0$, 1, or 2) in the decay of a pseudo-scalar particle into two vector particles ($B\to VV$).  A measurement of their relative contributions provides useful information to the puzzling scenario of $B\to VV$ polarizations. Theory predicts the transverse polarization fraction to be small, $f_T  = \mathcal{O}(m^2_V/m^2_B)\approx 4\%$, where $m_V~(m_B)$ is the mass  of the vector ($B$) meson, but measurements of $f_T \approx 50\%$ in $B^{0(+)} \to \phi K^{*(+)}$ decays disfavor this hierarchy \cite{B-fact}. Several \emph{ad hoc} solutions may accommodate the discrepancy without invoking non-SM physics. These include enhanced annihilation contributions, presence of transversely-polarized gluons, effects of electro-magnetic or charming penguins, long-distance re-scattering and others \cite{polarization1}. However, all these explanations are either model-dependent or non-conclusive, and the new physics option,  \eg\  $1+\gamma^5$  terms in the amplitude from scalar interactions or super-symmetric particles, remains valid. Polarization of the $\bs\to\phi\phi$ is considered crucial to discriminate among models. Ref.\cite{polarization2}, for instance, proposes to combine it with SU(3) symmetry to constrain the role of penguin annihilation amplitudes. The analysis of polarizations is in progress at CDF and $\mathcal{O}(5\%)$ resolutions on amplitudes are expected.
\section{Analysis of \bn\ and \lambdab\ decays into pairs of charmless hadrons}
Two-body decays mediated by the $b\to u$ quark-level transition have amplitudes sensitive to $\gamma$ (or $\phi_3)$, the least known angle of the CKM matrix; significant contributions from `penguin' transitions probe the presence of non--standard-model physics in loops;  in addition, the variety of open channels with similar final states allow cancellation of many common systematic effects and provides information to improve effective models of low-energy QCD. Analysis of these decays have a key role in the CDF flavor program, providing unique access to charmless \bs\  and \lambdab\ decays, and measurements in the \bd\ sector competitive with the $B$--factories \cite{bhh1}\cite{bhh2}. We briefly overview here the results based on $1~\lumifb$ of data collected by a dedicated displaced-track trigger similar to the one described in the previous section.\par
\begin{figure}[!h]
\begin{center}
\includegraphics[bb=0 115 611 630, width=0.375\textwidth, clip=true]{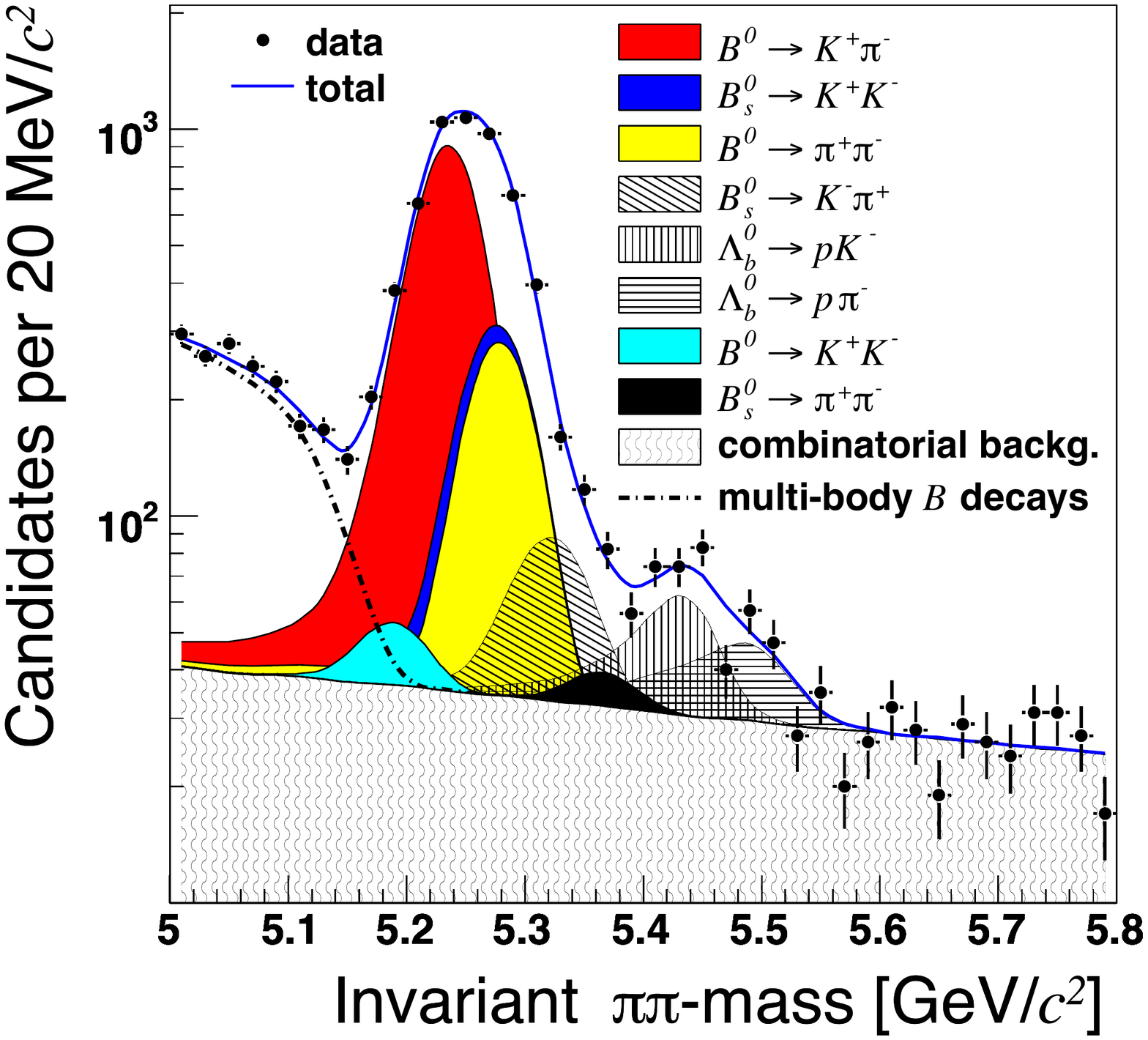}
\includegraphics[bb=0 115 611 630, width=0.375\textwidth, clip=true]{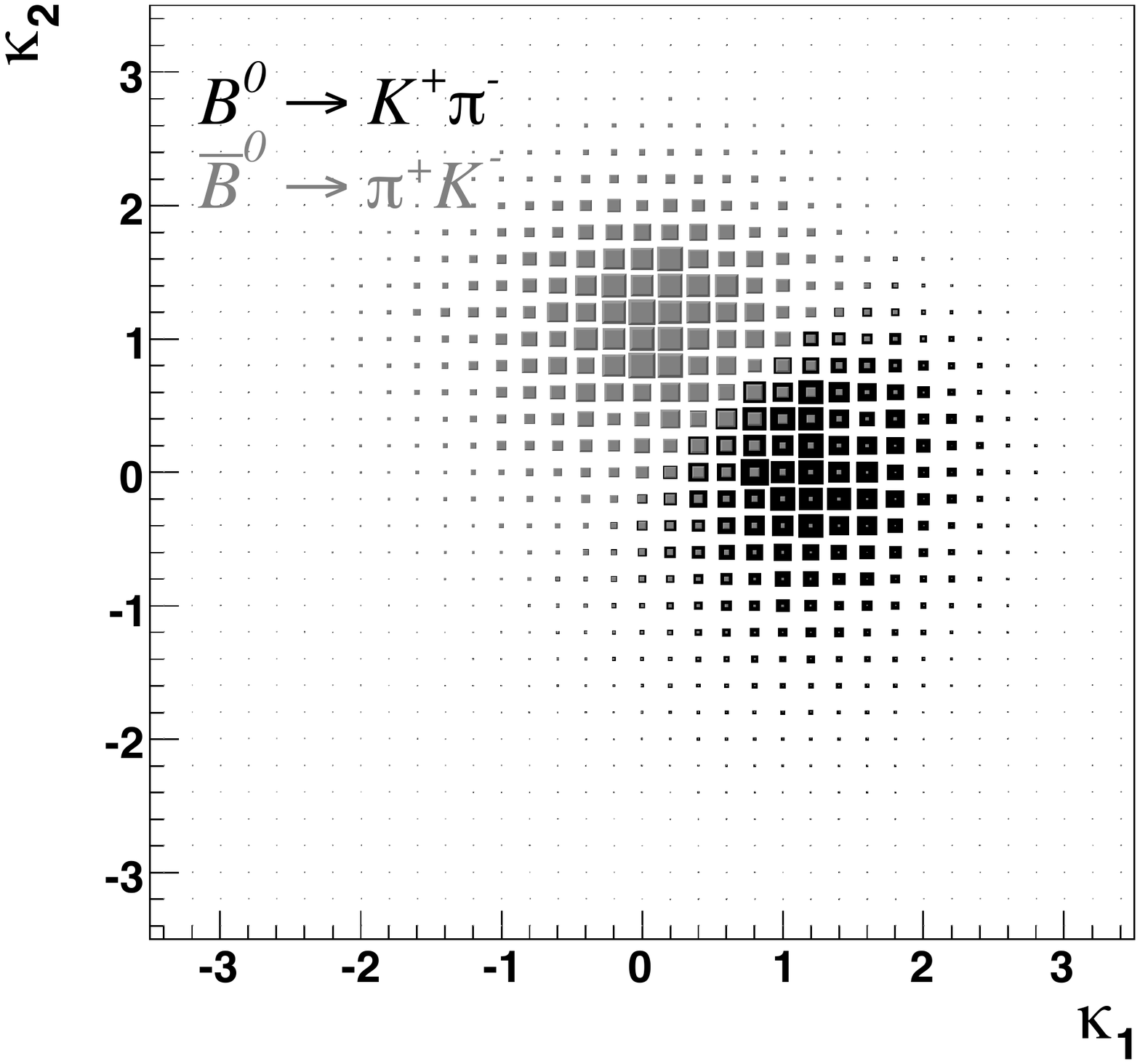}
\end{center}
\caption{Distribution of $\pi\pi$--mass  with fit projections overlaid (left). Separation between $K^+\pi^-$ and $K^-\pi^+$ states in the space of the \dedx\ of the two particles (right). The quantity $\kappa$ is the observed \dedx\ around the average pion response divided by the difference between average kaon and average pion responses.}
\label{fig:bhh}
\end{figure}

 An optimized offline selection isolates a clear signal of  a few thousands events over a smooth background dominated by combinatorial contributions (\fig{bhh}, left). 
Absence of event-specific particle identification and limited mass resolution  (approximately 22~\massmev\ in these decays) cause several channels to appear overlapping in a single peak. A five-dimensional likelihood fit relying on kinematics differences between decays (correlation between the two-body mass with arbitrary mass assignment to the final state particles and their momentum imbalance), and particle identification from the measurement of specific ionization in the drift chamber  ($1.5\sigma$ kaon-pion separation from \dedx, \fig{bhh}, right) allow statistical determination of the individual contributions. 
\begin{table}[!h]
\centering
\begin{tabular}{lccrclcc}
\hline\hline
Decay mode    &&  \multicolumn{3}{c}{Relative branching fraction (\br)}   && Absolute \br ($10^{-6}$)\\
\hline
\bdpipi\	&& \fracpipisukpi\    &=&    $0.259 \pm 0.017 \pm 0.016$  && $5.02 \pm 0.33 \pm 0.35$   \\
\bskk	\	&&   \frackksukpi\      &=&    $0.347 \pm 0.020 \pm 0.021$  && $24.4 \pm 1.4 \pm 3.5$      \\
\bskpi\	&&   \fracskpisukpi\   &=&    $0.071 \pm 0.010 \pm 0.007$  && $5.0 \pm 0.7 \pm 0.8$      \\
\bspipi\	&&   \fracspipisukpi\   &=&   $0.007 \pm 0.004 \pm 0.005$  && $<1.2$ at 90\% C.L. 		\\
\bdkk\	&&   \fracdkksukpi\   &=&     $0.020 \pm 0.008 \pm 0.006$  && $<0.7$ at 90\% C.L. 		\\
\lambdabpk\ &&   \fracpksukpi\   &=&     $0.066 \pm 0.009 \pm 0.008$  && 	$5.6 \pm 0.8 \pm 1.5$  	\\
\lambdabppi\ &&   \fracppisukpi\   &=&     $0.042 \pm 0.007 \pm 0.006$  && 	$3.5 \pm 0.6 \pm 0.9$  	\\
\hline\hline
\end{tabular}
\caption{Measured branching fractions. Absolute results were derived by normalizing to the current known value $\br(\bdkpi) = (19.4\pm 0.6) \times 10^{-6}$, and assuming $f_{s}/f_{d}= 0.276\pm0.034$ and $f_{L}/f_{d}= 0.230\pm0.052$ for the production fractions \cita{pdg}. The first quoted uncertainty is statistical, the second is systematic.}
\label{tab:BRbhh}
\end{table}
Large $D^0 \to K^-\pi^+$ samples are used both to model accurately mass line-shapes and for \dedx\ calibration. Event fractions for each channel are corrected for trigger and selection efficiencies (from simulation and data) to extract the decay-rates (tab.~\ref{tab:BRbhh}).
We report the first observation of the decays \bskpi, \lambdabpk, and \lambdabppi, and world-leading measurements of (upper limits on) the \bskk\  (\bspipi) branching fractions \cite{bhh2}. The measured ratio $\br(\lambdabppi)/\br(\lambdabpk) = 0.66 \pm 0.14\stat \pm 0.08\syst$ shows that significant penguin contributions compensate the Cabibbo (and kinematic) suppression expected at tree level. We also report first measurements of the direct \CP-violating asymmetries $\acp(\bskpi)=[39\pm 15\stat\pm 8\syst]\%$, $\acp(\lambdabppi)=[3\pm17\stat\pm5\syst]\%$, and $\acp(\lambdabpk)=[37\pm17\stat\pm3\syst]\%$. Direct \CP\ violation in the \bdkpi\ mode is measured as $\acp(\bdkpi)=[-8.6\pm2.3\stat\pm0.9\syst]\%$, which is consistent and close to be competitive with the final $B$--factories results.
The analysis of the $5~\lumifb$ sample is now in progress. We expect to reconstruct more than 15,000 \bhh\ decays,  where new modes could be observed and \CP-violating asymmetries measured with doubled precision.
\section{Concluding remarks and outlook}
CDF keeps harvesting from its rich and unique program on charmless $B$ decays in hadron collisions, pioneered since year 2002.
The update of the $\bs\to\phi\phi$ analysis to 2.9~\lumifb\ is the latest addition. The 37\% uncertainty on the branching fraction is almost half of the previous determination, and is dominated by the uncertainty on \br(\bsjpsiphi), which may soon reduced by the \belle\ experiment. The polarization analysis is in progress with expected uncertainties of 5\%. In the update to 1~\lumifb of the two-body analysis, the decays \bskpi, \lambdabppi, and  \lambdabpk\ were newly observed and their branching fractions and \CP-violating asymmetries measured. 
The observed hierarchy $\br(\lambdabpk) > \br(\lambdabppi)$ is inconsistent with recent theory predictions \cite{Lb}, suggesting significant contributions from penguin amplitudes in hadronic \lambdab\ decays.\par This fruitful program is reaching its maturity, but it's far from being exhausted. By October 2010 CDF will have more than 8~\lumifb\ of \emph{physics quality} data on permanent storage, which may reach $10~\lumifb$ after one year, if Tevatron Run II will be further extended. These 2.5--10 factors in increase of sample-size over analyses shown here set the scale of what's coming next. Not only current analyses, mostly limited by statistical uncertainties and by systematic effects that scale with statistics, will be improved.  Avenues for new measurements are opening-up: an accurate determination of the polarization structure of $\bs\to\phi\phi$ decays will contribute useful information on the puzzling scenario in $B\to VV$ decays; search for a large mixing phase in penguin-dominated $\bs\to\phi\phi$ decays will supplement the analogous effort in tree-dominated \bsjpsiphi\ modes \cita{sin2betas}; comparison of  \bd\ and \bs\ decay rates in common $K^+\pi^-$ final states will provide stringent model-independent constraints on non-SM physics contributions \cita{lipkin}, exploration of time-dependent \CP-violating asymmetries in \bhh\ decays will increase our knowledge of the angle $\gamma$ ($\phi_3$).\par While the first, exciting whimpers of LHC are emerging these days, CDF sits on a goldmine of data and a few exciting years of competition with LHCb are coming.
 \acknowledgments
 Thanks to the CDF and Tevatron people for producing these results and to the conference organizers for having patiently waited my long overdue contribution.
  
\end{document}